\documentclass[prb,aps]{revtex4}

\usepackage[dvips]{graphics}
\usepackage{amssymb}
\usepackage{psfig}

\begin{document}


\title{ Josephson effect in double-barrier
superconductor-ferromagnet junctions }
\author{Z. Radovi\'c,$\ ^{1}$ N. Lazarides,$\ ^{2}$ and N. Flytzanis$\ ^{3}$
}
\address{
$\ ^{1}$Department of Physics, University of Belgrade, P.O. Box
368,
11001 Belgrade, Yugoslavia  \\
$\ ^{2}$ Department of Materials Science and Technology,
University of Crete, P.O. Box 2208, 71003 Heraklion, Greece \\
$\ ^{3}$ Department of Physics, University of Crete, P.O. Box
2208, 71003 Heraklion, Greece }

\begin{abstract}
We study the Josephson effect in ballistic double-barrier SIFIS
planar junctions, consisting of bulk superconductors (S), a clean
metallic ferromagnet (F), and insulating interfaces (I). We solve
the scattering problem based on the Bogoliubov--de Gennes
equations and derive a general expression for the dc Josephson
current, valid for arbitrary interfacial transparency and Fermi
wave vectors mismatch (FWVM). We consider the coherent regime in
which quasiparticle transmission resonances contribute
significantly to the Andreev process. The Josephson current is
calculated for various parameters of the junction, and the
influence of both interfacial transparency and FWVM is analyzed.
For thin layers of strong ferromagnet and finite interfacial
transparency, we find that coherent (geometrical) oscillations of
the maximum Josephson current are superimposed on the oscillations
related to the crossovers between 0 and $\pi$ states. For the same
case we find that the temperature-induced $0-\pi$ transition
occurs if the junction is very close to the crossover at zero
temperature.
\end{abstract}


\maketitle

\section{Introduction}

Proximity effects in superconductor (S) --ferromagnet (F) hybrid
structures have been studied for some time already.\cite{tedrow}
Recent realization of a $\pi$ state in metallic SFS junctions,
\cite{ryazanov,ryazanov1,kontos} has reinvigorated interest in
further experimental and theoretical
studies.\cite{geers,bourgeois,kontos1,fogelstrom,barash,chtchelkatchev,krivoruchko,bergeret,golubov,barash1,zyuzin,zr,golubov1,heikkila,halterman,zareyan,baladie,bergeret1}
Nowadays, understanding of  the coherent geometrical effects  in
ballistic heterojunctions is also becoming more  important,
\cite{brinkman,ingerman,kikuchi,bozovic} due to the progress in
nanofabrication technology and the improvement of experimental
techniques.\cite{nevirkovets,schulze,moussy}

The possibility of a $\pi$ state in superconductors coupled
through a magnetically active material (an insulating barrier
containing paramagnetic impurities, or a ferromagnetic metal) has
been proposed long ago. \cite{bulaevskii,buzdin} In  the $\pi$
state of an SFS structure, in contrast to the usual (0) state, the
phase shift equal to $\pi$ across the junction in the ground state
reverses the direction of the supercurrent flow,\cite{ryazanov}
and drastically changes the density of states (DOS) in F
metal.\cite{kontos} Following the theoretical
prediction,\cite{radovic,buzdin1} an evidence for $\pi$ states in
proximity-coupled  S-F superlattices has been sought previously in
the oscillations of the superconducting critical temperature $T_c$
as a function of the F-layers thickness.\cite{jiang,obi} More
recently, $\pi$ states have been observed in nonmagnetic junctions
of high-$T_c$ superconductors\cite{vanharlingen} and in
out-of-equilibrium mesoscopic superconducting
structures.\cite{baselmans}

Oscillations  of the maximum Josephson current $I_c$ and of the
local DOS, with thickness and strength of the F layer, are
prominent features of SFS metallic junctions. These oscillations
are related to the crossovers between 0 and $\pi$ states, the
$I_c$ minima being located at the crossover points.\cite{kontos1}
Nonmonotonic temperature variation of $I_c$, also related to the
transition from $\pi$ to 0 state, has been observed
recently.\cite{ryazanov} This effect is studied theoretically for
superconducting junctions with different barriers, such as
magnetically active insulating interfaces,\cite{fogelstrom,barash}
metallic FIF layers (including insulating inhomogeneity and
nonuniform
magnetization),\cite{chtchelkatchev,krivoruchko,bergeret,golubov,barash1}
and ferromagnetic-metal barriers with mesoscopic
disorder.\cite{zyuzin} The temperature induced $0 - \pi$
transition is attributed to the spin discrimination of Andreev
bound states in the case of finite transparency and strong
ferromagnetic influence.\cite{barash}

Another characteristic feature is a strong contribution of higher
harmonics to the current-phase relation $I(\phi)$ in the vicinity
of the crossover points, as has been shown for metallic SFS
junctions in both clean and diffusive limits,\cite{zr,golubov1}
for Josephson junctions with magnetically active insulating
interface,\cite{barash,Tanaka 97} and for non-equilibrium
supercurrent through mesoscopic ferromagnetic weak
links.\cite{heikkila} This implies that the energy of the junction
in the vicinity of the crossover has two minima as a function of
$\phi$, at $\phi=0$ and $\phi=\pi$. The resulting coexistence of
stable and metastable 0 and $\pi$ states in the crossover region
can generate two flux jumps per one external flux quantum in
SQUIDs.\cite{zr}

The underlying microscopic mechanism is well understood. The
Andreev process, recognized as the mechanism of
normal-to-supercurrent conversion, \cite{andreev,BTK,Furusaki
Tsukada,beenakker} is modified at F-S interfaces due to the spin
imbalance in the ferromagnet.\cite{dejong} As a result, the
superconducting pair amplitude induced in F by the proximity to S
is spatially modulated.\cite{halterman} The current-carrying
Andreev bound states are split and shifted in an oscillatory way
under the influence of the ferromagnet.\cite{zareyan} The
crossovers between 0 and $\pi$ states and highly nonsinusoidal
current-phase relation follow from the strong spin polarization of
the Andreev states.\cite{barash,Tanaka 97}

Several quantities characterizing the proximity effect, such as
$I(\phi)$ and local DOS, have been studied in a number of
theoretical works for different geometries of S-F structures,
using the quasiclassical approach in both clean and diffusive
limits.\cite{bergeret,golubov,zareyan,baladie,bergeret1} However,
a theory of the phase-coherent electronic transport in mesoscopic
structures should be based on the solutions of Gor'kov or
Bogoliubov--de Gennes (BdG) equations. The coherence effects have
been studied recently for double-barrier SINIS junctions
containing an interlayer of a clean nonmagnetic metal (N) with
insulating interfaces (I),\cite{brinkman} and for FISIF
junctions.\cite{kikuchi,bozovic}

In this paper, we study the simultaneous influence of two
insulating barriers on the supercurrent flow in ballistic
double-barrier SIFIS planar junctions. The influence of different
band-widths in two metals (the Fermi wave vector  mismatch --
FWVM) is also included. We limit ourselves to conventional
($s$-wave) superconductors. Assuming  a constant pair potential in
the S electrodes, we solve analytically the scattering problem
based on the BdG equations, and derive a  general expression for
the Josephson current. This approach has been applied previously
to SIS and  FIS junctions with magnetically active insulating
interface, both for the $s$-wave and for an unconventional pairing
symmetry in superconductors. \cite{Tanaka 97,Tanaka 10,beasley} In
a  limiting case, our expression gives a generalization of the
previous formulae for the Josephson current,\cite{Furusaki Tsukada
91,arnold} that includes the finite interfacial  transparency in
SINIS junctions. Strong geometrical oscillations of $I_c$ in the
junctions with thin normal-metal interlayers and finite
interfacial transparency are related to the contribution of
quasiparticle transmission resonances.\cite{zetp} For thin layers
of a strong ferromagnet, these oscillations are superimposed on
the oscillations related to the crossovers between 0 and  $\pi$
states. Lower transparency and FWVM shift the crossover points and
narrow the adjacent regions of coexisting 0 and $\pi$ states with
highly nonsinusoidal current-phase relation. In a junction with
finite transparency, with or without FWVM, and with strong
ferromagnetic influence, the temperature-induced transition
between $0$ and $\pi$ states occurs if the junction is
sufficiently close to the  crossover at zero temperature. In that
case, the transition region of coexisting 0 and $\pi$ states is
considerably large.

The paper is organized as follows. In Sec. II, the scattering
problem based on the BdG equations is solved analytically for a
planar SIFIS junction, and general expressions for both the
Andreev and the normal reflection amplitudes are presented. In
Sec. III, an expression for the Josephson current is derived; this
section includes an analysis of the influence of the junction
parameters on the crossovers between 0 and $\pi$ states and on the
current-phase relation. Concluding remarks are given in Sec. IV.

\section{The scattering problem
} We consider the following model for a  planar double-barrier
SIFIS junction: a  ferromagnetic  layer of thickness $d$ is
connected to superconductors by insulating nonmagnetic interfaces.
We assume that both metals are clean, that the left (L) and the
right (R) superconductors are equal, and so are the interface
barriers. We use the Stoner model for the ferromagnet (a uniform
magnetization is parallel to the layers), and describe the
quasiparticle propagation by the Bogoliubov--de Gennes equation
\begin{eqnarray}
\left(
\begin{array}{ccc}
  H_0({\bf r})-\rho_{\sigma}h({\bf r}) && \Delta({\bf r}) \\
  \Delta^{*}({\bf r}) && -H_0({\bf r})+\rho_{\bar{\sigma}}h({\bf
r})
\end{array}
\right) \left(
\begin{array}{c}
    u_{\sigma}({\bf r}) \\
    v_{\bar{\sigma}}({\bf r})
  \end{array}
  \right)
=~E \left(
\begin{array}{c}
    u_{\sigma}({\bf r}) \\
    v_{\bar{\sigma}}({\bf r})
  \end{array}
  \right)  .
 \label{BdG}
\end{eqnarray}
Here, $H_{0}({\bf r})=-\hbar^{2}\nabla^{2}/2m+W({\bf r})+U({\bf
r})-\mu$, where $U({\bf r})$ and $\mu$ are the electrostatic and
the  chemical potential, respectively. The interface potential is
modelled  by $W({\bf r})=\hat{W}[\delta(z)+\delta(z-d)]$, where
the $z$ axis is perpendicular to the layers,  and $\delta(z)$ is
the Dirac $\delta$-function. In Eq. (\ref{BdG}), $\sigma$ denotes
the spin orientation ($\sigma =\uparrow ,\downarrow$),
$\bar{\sigma}$ is opposite to $\sigma$, $E$ is the quasiparticle
energy with respect to $\mu$, $h({\bf r}) = h\Theta(z)\Theta(d-z)$
is the exchange potential, where $\Theta(z)$ is the Heaviside step
function, and $\rho_{\sigma}$ is 1 ($-1$) for  $\sigma =\uparrow
(\downarrow)$. Neglecting the self-consistency of the
superconducting pair potential, $\Delta({\bf r})$ is taken in the
form
\begin{eqnarray}
    \Delta({\bf r}) = \Delta  \left[ e^{i \phi_{L}} \Theta(-z)
      + e^{i \phi_R}  \Theta(z-d) \right]  ,
\end{eqnarray}
where $\Delta$ is the  bulk superconducting gap, and $\phi =
\phi_R - \phi_L$ is the macroscopic phase difference across  the
junction. The temperature dependence of $\Delta$ is given by
$\Delta (T) = \Delta (0) \tanh\left( 1.74\sqrt{T_c/T - 1}
\right)$. The electron effective mass $m$ is assumed to be the
same for both metals, $\mu-U({\bf r})$ is the Fermi energy of the
superconductor, $E^{(S)}_F$, or the mean Fermi energy of the
ferromagnet, $E^{(F)}_F=(E^\uparrow_F+E^\downarrow_F)/2$. Moduli
of the Fermi wave vectors, $k^{(S)}_F=\sqrt{2mE^{(S)}_F/\hbar^2}$
and $k^{(F)}_F= \sqrt{2mE^{(F)}_F/\hbar^2}$, may be different in
general, and in the following the FWVM will be parameterized by
$\kappa=k^{(F)}_F/k^{(S)}_F$.

The parallel component of the wave vector ${\bf k}_{||}$ is
conserved, and the wave function
\begin{equation}
\left(
\begin{array}{c}
    u_{\sigma}({\bf r}) \\
    v_{\bar{\sigma}}({\bf r})
  \end{array}
  \right)
  = \exp(i{\bf k}_{||} \cdot {\bf r} )
  \psi(z)
\end{equation}
satisfies  appropriate boundary conditions
\begin{eqnarray}
\label{bc1}
\psi(z)|_{z=0_-}&=&\psi(z)|_{z=0_+},\\
\frac{d\psi(z)}{dz}\Big|_{z=0_-}&=&\frac{d\psi
(z)}{dz}\Big|_{z=0_+} -\frac{2m\hat{W}}{\hbar ^2}\psi
(0),\\
\psi(z)|_{z=d_-}&=&\psi(z)|_{z=d_+},\\
\frac{d\psi(z)}{dz}\Big|_{z=d_-}&=&\frac{d\psi
(z)}{dz}\Big|_{z=d_+}-\frac{2m\hat{W}}{\hbar ^2}\psi(d)
\label{bc4}.
\end{eqnarray}
The four independent solutions of Eq. (\ref{BdG}) correspond to
four types of quasiparticle injection processes: an electron-like
quasiparticle (ELQ) or a hole-like quasiparticle (HLQ) injected
from either the left or from the right superconducting electrode
(see Fig. 2 in Ref. \cite{Furusaki Tsukada}).

For the injection of an ELQ  from the left superconductor  with
energy $E > \Delta$  and angle of incidence $\theta$ (measured
from the $z$-axis),  $\psi (z)$ has the following form
\begin{eqnarray}
\label{psiL} \psi_1 (z)=
  \left\{
  \begin{array}{ll}
     [\exp(ik^+ z) + b_1 \exp(-ik^+z) ]
    \left(
    \begin{array}{c}
         \bar{u} e^{ i \phi_L / 2}   \\
         \bar{v} e^{-i \phi_L / 2}
   \end{array}
   \right)
    +  a_1 \exp(ik^- z)
   \left(
   \begin{array}{c}
         \bar{v}  e^{ i \phi_L / 2} \\
         \bar{u}  e^{-i \phi_L / 2}
     \end{array}
    \right)  & z < 0,   \\
    \\
 \left[ C_1 \exp( iq^+_{\sigma} z)
     + C_2 \exp(-iq^+_{\sigma} z)
     \right]
  \left(
  \begin{array}{c}
    1 \\
    0
  \end{array}
  \right)
   + [C_3 \exp(iq^-_{\bar{\sigma}}z) + C_4 \exp(-iq^-_{\bar{\sigma}}z)]
\left(
\begin{array}{c}
    0 \\
    1
\end{array}
\right)  & 0<z<d,  \\
  \\
  c_1 \exp(ik^+ z)
  \left(
  \begin{array}{c}
   \bar{u}  e^{i \phi_R / 2}   \\
   \bar{v}  e^{- i \phi_R / 2}
     \end{array}
   \right)
   +  d_1 \exp(-ik^- z)
   \left(
   \begin{array}{c}
     \bar{v} e^{i \phi_R / 2}  \\
     \bar{u} e^{- i \phi_R / 2}
   \end{array}
   \right) & z>d .
   \\
   \end{array}
   \right.
\end{eqnarray}
Here, $\bar{u} =  \sqrt{(1+ \Omega / E) / 2}$ and $\bar{v} =
\sqrt{(1- \Omega / E) / 2}$ are the BCS amplitudes, and
$\Omega=\sqrt{E^2-\Delta^2}$. Perpendicular ($z$) components of
the wave vectors are
\begin{eqnarray}
 \label{kz}
  k^\pm = \left[ (2m/\hbar ^2)(E^{(S)}_F\pm\Omega)-{\bf k}^2_{||} \right]^{1/2},
\end{eqnarray}
and
\begin{eqnarray}
 \label{qz}
   q^\pm_{\sigma}
      =\left[ (2m/\hbar^2)(E^{(F)}_F+\rho_{\sigma}h \pm E)-{\bf k}^2_{||} \right]^{1/2},
\end{eqnarray}
where $|{\bf k}_{||}|=\sqrt{(2m/\hbar^2)(E^{(S)}_F +
\Omega)}~\sin\theta$ is  the conserved parallel component. The
coefficients $a_1$, $b_1$, $c_1$, and $d_1$ are, respectively, the
probability amplitudes of the generalized  Andreev reflection as a
HLQ, normal  reflection  as an ELQ, transmission to the right
electrode as an ELQ, and  transmission to the right  electrode as
an  HLQ.\cite{Furusaki Tsukada} Amplitudes of electrons and holes
propagating in the ferromagnetic layer are given by the
coefficients $C_1$ through $C_4$. All amplitudes are
$\sigma$-dependent through the Zeeman terms in $q^\pm_{\sigma}$,
but  there is no spin current across the junction in contrast to
FIS and FISIF geometry.\cite{dejong,beasley,bozovic} ELQ and the
Andreev-reflected HLQ have identical spin orientations in absence
of spin-flip processes and for singlet-state pairing in SIFIS
geometry.\cite{barash1}

For the injection of a HLQ  from the left superconductor with
energy $E > \Delta$ and angle of incidence $\theta$, $\psi (z)$ is
\begin{eqnarray}
\label{psiL2} \psi_2 (z) =
   \left\{
    \begin{array}{ll}
  [\exp(-ik^- z) + b_2 \exp(ik^-z) ]
\left(\begin{array}{c}
     \bar{v} e^{ i \phi_L / 2}   \\
     \bar{u} e^{-i \phi_L / 2}
   \end{array}
   \right)
    + a_2 \exp(-i k^+ z)
    \left(
    \begin{array}{c}
      \bar{u}  e^{ i \phi_L / 2} \\
      \bar{v}  e^{-i \phi_L / 2}
    \end{array}
    \right) & z<0 , \\
    \\
    \left[ C_1' \exp(iq^+_{\sigma}z)
         + C_2' \exp(-iq^+_{\sigma}z)
     \right]
\left(\begin{array}{c}
    1 \\
    0
  \end{array}\right)
    + [ C_{3}' \exp(iq^-_{\bar{\sigma}}z) + C_{4}' \exp(-iq^-_{\bar{\sigma}}z) ]
\left(\begin{array}{c}
    0 \\
    1
  \end{array}
  \right)  & 0<z<d \\
  \\
     c_2 \exp(-i k^- z)\left(
 \begin{array}{c}
     \bar{v}  e^{ i \phi_R / 2}   \\
     \bar{u}  e^{-i \phi_R / 2}
   \end{array}\right)
    + d_2 \exp( ik^+ z)
    \left(\begin{array}{c}
      \bar{u} e^{ i \phi_R / 2}  \\
      \bar{v} e^{-i \phi_R / 2}
    \end{array}\right)  & z>d , \\
\end{array}
\right.
\end{eqnarray}
where the  wave vectors are given by Eqs. (\ref{kz}) and
(\ref{qz}), with $|{\bf k}_{||}|=\sqrt{(2m/\hbar^2)(E^{(S)}_F -
\Omega)}~\sin\theta$. Analogously, one can  write  $\psi_3$ and
$\psi_4$ for an injection of ELQ and HLQ from the right
superconductor, respectively.

Solutions of Eqs. (\ref{bc1})--(\ref{bc4}) for the scattering
amplitudes  can be  simplified significantly if one neglects,
except in exponentials, small terms $\Omega / E^{(S)}_F \ll 1$ and
$E /  E^{(F)}_F \ll 1$  in the wave vectors. In the following, $|
{\bf k}_{||} | = k^{(S)}_F \sin\theta$, the wave vectors $k^\pm$
are replaced by $k= \sqrt{  {k_F^{(S)}}^2 - {\bf k}_{||}^2 } =
k^{(S)}_F \cos\theta$, and  in the pre-exponential factors only,
$q_\sigma^+$ and $q_{\bar{\sigma}}^-$ are  replaced by $q_\sigma$
and  $q_{\bar{\sigma}}$, where $q_\sigma =
   \sqrt{ {k_F^{(F)}}^2 ( 1 + \rho_\sigma h /  E^{(F)}_F )
  -{\bf k}^2_{||} }$.
Physically important  oscillations of the scattering amplitudes,
both  rapid and slow on the atomic scale $1 / k_F^{(F)}$, are
characterized respectively by the  exponents
\begin{equation}
\label{zeta}
    \zeta_\sigma^\pm =d \left(q^+_{\sigma}\pm  q^-_{\bar{\sigma}} \right).
\end{equation}
First  we present the  results for  $a_1$ and $b_1$, given by
\begin{eqnarray}
  \label{a1}
   a_1
    = \frac{2\Delta}{G} \left[
    -4 \tilde{q}_{\sigma} \tilde{q}_{\bar{\sigma}}
              \left( E \cos\phi + i \Omega \sin\phi \right)
    + {\cal A}_1^- \cos(\zeta_\sigma^-) - {\cal A}_1^+  \cos(\zeta_\sigma^+)
    - i {\cal A}_2^- \sin(\zeta_\sigma^-) + i {\cal A}_2^+  \sin(\zeta_\sigma^+)
    \right] ,
\end{eqnarray}
and
\begin{eqnarray}
  \label{b1}
   b_1  = \frac{2\Omega}{G} \left[
            {\cal B}_1^-  \cos(\zeta_\sigma^-) - {\cal B}_1^+  \cos(\zeta_\sigma^+)
       - i {\cal B}_2^- \sin(\zeta_\sigma^-) + i {\cal B}_2^+  \sin(\zeta_\sigma^+)
       \right] ,
\end{eqnarray}
where
\begin{eqnarray}
     {\cal A}_1^{\pm}  &=&
     \left( \tilde{q}_\sigma
        \mp  \tilde{q}_{\bar{\sigma}} \right)
      \left[
      E \left( \tilde{q}_\sigma
       \mp  \tilde{q}_{\bar{\sigma}} \right)  -
       i \Omega {Z_\theta}
        \left( \tilde{q}_\sigma
    \pm \tilde{q}_{\bar{\sigma}} \right)
    \right] ,
       \nonumber \\
    {\cal A}_2^{\pm} &=&
       \Omega
       \left( \tilde{q}_\sigma  \mp \tilde{q}_{\bar{\sigma}}
        \right)
     \left[ 1 + Z_\theta^2
      \mp \tilde{q}_\sigma
      \tilde{q}_{\bar{\sigma}} \right] ,
      \label{as}
\end{eqnarray}
and
\begin{eqnarray}
     {\cal B}_1^{\pm} &=&
       E  \left( 1  - i Z_\theta \right)
        \left[ (\tilde{q}_\sigma)^2 - (\tilde{q}_{\bar{\sigma}})^2
    \right]
       \nonumber  \\
       &-& \Omega \left\{
           \left( 1 + Z_\theta^2 \right)
       \left( 1 - i Z_\theta \right)^2
          - (\tilde{q}_\sigma)^2  (\tilde{q}_{\bar{\sigma}})^2
          + i Z_\theta  \left( 1 - i Z_\theta \right)
      \left[ (\tilde{q}_{\sigma})^2
      \pm  4\tilde{q}_{\sigma}
      \tilde{q}_{\bar{\sigma}} +
     (\tilde{q}_{\bar{\sigma}})^2
      \right]  \right\},
    \nonumber  \\
    {\cal B}_2^{\pm}  &=&
    - E \left( \tilde{q}_\sigma
        \mp \tilde{q}_{\bar{\sigma}} \right)
     \left[
     \left( 1 - i Z_\theta \right)^2
     \pm \tilde{q}_\sigma \tilde{q}_{\bar{\sigma}}
     \right]
     \nonumber  \\
    &+& \Omega
    \left( \tilde{q}_\sigma \pm \tilde{q}_{\bar{\sigma}} \right)
     \left[
     \left( 1 + 2 i Z_\theta \right) \left( 1 - i Z_\theta \right)^2
      \mp  ( 1 - 2 i Z_\theta )
      \tilde{q}_\sigma \tilde{q}_{\bar{\sigma}}
   \right].
      \label{bs}
\end{eqnarray}
The common denominator is given by
\begin{eqnarray}
  \label{gamma}
  G = 8\Delta^2  \tilde{q}_\sigma \tilde{q}_{\bar{\sigma}} \cos\phi
     - {\cal G}_1^- \cos(\zeta_\sigma^-) + {\cal G}_1^+ \cos(\zeta_\sigma^+)
     + i {\cal G}_2^- \sin(\zeta_\sigma^-) - i {\cal G}_2^+ \sin(\zeta_\sigma^+)  ,
\end{eqnarray}
where
\begin{eqnarray}
   {\cal G}_1^{\pm} &=&
     \left\{ E
      \left( \tilde{q}_{\sigma} \mp \tilde{q}_{\bar{\sigma}} \right)
       + \Omega \left[ 1 + Z_\theta^2
      - i Z_\theta
      \left( \tilde{q}_{\sigma} \pm \tilde{q}_{\bar{\sigma}} \right)
      \mp  \tilde{q}_{\sigma} \tilde{q}_{\bar{\sigma}}
      \right]
       \right\}^2
    \nonumber \\
      &+& \left\{ E\left( \tilde{q}_\sigma \mp \tilde{q}_{\bar{\sigma}}
     \right)
      - \Omega \left[ 1 + Z_\theta^2
    + i Z_\theta
     \left( \tilde{q}_\sigma \pm  \tilde{q}_{\bar{\sigma}} \right)
    \mp  \tilde{q}_\sigma \tilde{q}_{\bar{\sigma}}
     \right]
    \right\}^2,
    \nonumber \\
   {\cal G}_2^{\pm}  &=&
    4 \Omega
    \left( 1 + Z_\theta^2
    \mp \tilde{q}_\sigma
    \tilde{q}_{\bar{\sigma}} \right)
    \left[
    E\left( \tilde{q}_\sigma \mp \tilde{q}_{\bar{\sigma}} \right)  -
     i \Omega Z_\theta
      \left( \tilde{q}_\sigma \pm \tilde{q}_{\bar{\sigma}} \right)
     \right]  .
      \label{gs}
\end{eqnarray}
Here, we introduced  the normalized quantities $\tilde{q}_\sigma =
{q}_\sigma / k$, and $Z_\theta = Z / \cos\theta$, where
$Z={2m\hat{W}}/\hbar ^2 k^{(S)}_F$ is  the parameter measuring the
strength of each insulating interface. Note that all  amplitudes
are functions of  $E,~\sigma,~\theta$, and $\phi$, for given
$\Delta$, $h$, and  $Z$.

The Andreev reflection amplitudes $a_2$ and $a_1$  are  simply
connected. For the same $E$, $\sigma$ and $\theta$ in our
approximation we get
\begin{eqnarray}
 \label{a2}
   a_2 (\phi) = a_1 ( -\phi) ,
\end{eqnarray}
which  is in agreement with the  detailed balance relations.
\cite{Furusaki Tsukada} Expression for $b_2$  can be given in a
form  similar to  $b_1$, so that
\begin{eqnarray}
  \label{b2}
   b_2  = \frac{2\Omega}{G} \left[
            \bar{{\cal B}}_1^-  \cos(\zeta_\sigma^-)
        - \bar{{\cal B}}_1^+  \cos(\zeta_\sigma^+)
       + i \bar{{\cal B}}_2^- \sin(\zeta_\sigma^-)
       - i \bar{{\cal B}}_2^+  \sin(\zeta_\sigma^+)
       \right] ,
\end{eqnarray}
with  $G$ given  by Eq. (\ref{gamma}), and
\begin{eqnarray}
     \bar{{\cal B}}_1^{\pm} &=&
       - E  \left( 1  + i Z_\theta \right)
        \left[ (\tilde{q}_\sigma)^2 - (\tilde{q}_{\bar{\sigma}})^2
    \right]
       \nonumber  \\
       &-& \Omega \left\{
           \left( 1 + Z_\theta^2 \right)
       \left( 1 + i Z_\theta \right)^2
          - (\tilde{q}_\sigma)^2  (\tilde{q}_{\bar{\sigma}})^2
          - i Z_\theta  \left( 1 + i Z_\theta \right)
      \left[ (\tilde{q}_{\sigma})^2
      \pm  4\tilde{q}_{\sigma}
      \tilde{q}_{\bar{\sigma}} +
     (\tilde{q}_{\bar{\sigma}})^2
      \right]  \right\}  ,
    \nonumber  \\
    \bar{{\cal B}}_2^{\pm}  &=&
    - E \left( \tilde{q}_\sigma
        \mp \tilde{q}_{\bar{\sigma}} \right)
     \left[
     \left( 1 + i Z_\theta \right)^2
     \pm \tilde{q}_\sigma \tilde{q}_{\bar{\sigma}}
     \right]
     \nonumber  \\
    &-& \Omega
    \left( \tilde{q}_\sigma \pm \tilde{q}_{\bar{\sigma}} \right)
     \left[
     \left( 1 - 2 i Z_\theta \right) \left( 1 + i Z_\theta \right)^2
      \mp  ( 1 + 2 i Z_\theta )
      \tilde{q}_\sigma \tilde{q}_{\bar{\sigma}}
   \right] .
   \label{bs2}
\end{eqnarray}
Note that the normal-reflection amplitudes, $b_1$ through $b_4$,
are even functions of $\phi$. From the assumed  symmetry of the
junction, within our  approximation  it  follows that $a_3 =
a_2,~a_4 = a_1,~b_3 = b_1$, and $b_4 = b_2$.\cite{foot2}

In the corresponding NIFIN junction, when the superconductor
electrodes are in the normal state, the expressions for the normal
reflection amplitudes reduce to $b_1 = b_2 \equiv b_N$, where
\begin{eqnarray}
 \label{bn}
  b_N = \frac{
    2 Z_\theta \tilde{q}_\sigma \cos(d q_\sigma)
    + \left( 1 + Z_\theta^2 - \tilde{q}_\sigma^2 \right)
    \sin(d q_\sigma) }
    { 2 i ( 1 + i Z_\theta ) \tilde{q}_\sigma \cos(d q_\sigma)
    + \left( 1 + 2 i Z_\theta - Z_\theta^2 + \tilde{q}_\sigma^2
    \right)
    \sin(d q_\sigma) }.
\end{eqnarray}

Because of the conservation of ${\bf k}_{||}$, ELQ and HLQ undergo
the total reflection for $\theta > \theta_{c\sigma} =
\sin^{-1}\lambda_\sigma$, if $\lambda_\sigma = \kappa \sqrt{1 +
\rho_\sigma h/ E_F^{(F)} } < 1$. The corresponding $q_\sigma$
becomes imaginary and electrons and/or holes, depending on the
spin orientation, cannot propagate in the ferromagnetic layer.
However, the  contribution of an evanescent type of the  Andreev
reflection to the Josephson current is not
negligible,\cite{beasley} and should  be taken into account in the
finite geometry.

\section{The Josephson current}

The dc Josephson current at a given temperature can  be expressed
in terms of the Andreev reflection amplitudes by using the
temperature Green's function  formalism\cite{Furusaki Tsukada}
\begin{eqnarray}
  I = \frac{e \Delta}{2 \hbar} \sum_{\sigma,{\bf k}_{||}}
     k_B T \sum_{\omega_n}
       \frac{1}{2\Omega_n} (k_n^+ + k_n^-)
         \left( \frac{a_{n 1}}{k_n^+} - \frac{a_{n 2}}{k_n^-}
       \right) ,
\end{eqnarray}
where $k_n^+,~k_n^-$, and $a_{n1},~a_{n2}$ are obtained from
$k^+,~k^-$, and $a_1,~a_2$ by the analytic continuation $E
\rightarrow i \omega_n$, the Matsubara frequencies are $\omega_n =
\pi k_B T (2n+1)$ with  $n=0,\pm1, \pm2, ...$, and $\Omega_n =
\sqrt{\omega_n^2 + \Delta^2 }$. Performing integration over ${\bf
k}_{||}$ and using Eqs. (\ref{a1}) and (\ref{a2}), for the
Josephson current in a planar SIFIS junction we get
\begin{eqnarray}
  \label{main}
  I = \frac{4 \pi k_B T \Delta^2}{e R }
       \int_0^{\pi/2} d\theta \sin\theta \cos\theta
       \sum_{\omega_n,\sigma}
    \frac{ \tilde{q}_\sigma \tilde{q}_{\bar{\sigma}}
      \sin\phi }{G_n} .
\end{eqnarray}
Here,  $G_n$ is  $G$ given by Eq. (\ref{gamma}), with $E$ and
$\Omega$ replaced by $i \omega_n$ and $i \Omega_n$.  Note that $R=
2\pi^2 \hbar  /  S e^2 {k_F^{(F)}}^2$, where $S$ is   the area of
the junction, is the normal resistance only for $Z=0$, $\kappa=1$,
and $h=0$, when the normal reflection amplitude $b_N$ is equal to
zero. The resistance $R_N$ of the corresponding NIFIN junction can
be obtained from
\begin{eqnarray}
   \frac{R}{R_N} = \int_0^{\pi/2} d\theta \sin\theta \cos\theta
      \sum_{\sigma} \left( 1 - | b_N |^2 \right).
\end{eqnarray}

The spectrum of bound states in the interlayer is included in the
common denominator of the retarded Green's function. For
transparent nonmagnetic junctions without FWVM, when $R_N  /R =1$,
the condition $G(E)=0$ gives well-known phase-dependent and
spin-degenerate Andreev bound states with subgap
energies.\cite{Furusaki Tsukada 91} For resistive ferromagnetic
junctions, when $R_N /R > 1$, the spectrum of Andreev bound states
is modified by the coherent contribution of geometrical resonances
in the ferromagnet (described by the  rapidly oscillating terms)
and by the Zeeman splitting.

For  a  weak ferromagnet, $h/E_F^{(F)} \ll 1$, wave vectors
${q}_{\sigma}$ and ${q}_{\bar{\sigma}}$ can both be replaced by
${q} =  \sqrt{ {k_F^{(F)}}^2 - {\bf k_{||}}^2 }$, so that
$\tilde{q}_{\sigma}, \tilde{q}_{\bar{\sigma}} \rightarrow
\tilde{q} = q/k$. Also, $\zeta_\sigma^\pm$ can be approximated as
$\zeta_\sigma^- \simeq {2 d}
  \left( E + \rho_\sigma h \right) /{\hbar v}$ and
$\zeta_\sigma^+ \simeq 2 q d $, where $v =  \hbar  q / m $ is the
$z$ component of the Fermi velocity in the absence of FWVM. In
this limit, the general formula, Eq. (\ref{main}), reduces to
\begin{eqnarray}
  \label{mains}
  I = \frac{ \pi k_B T \Delta^2}{e R }
       \int_0^{\pi/2} d\theta \sin\theta \cos\theta
       \sum_{\omega_n} \frac{1}{2} \sum_{\sigma}
    \frac{\sin\phi }{\Gamma_n} ,
\end{eqnarray}
with
\begin{eqnarray}
  \label{gamman}
    \Gamma_n =
     &~& \Delta^2 \cos\phi
      + \left( K^2 \Omega_n^2 + \omega_n^2 \right)
   \cosh\left[ \frac{2(\omega_n - i \rho_\sigma h) d}{\hbar v} \right]
      + 2  K \omega_n \Omega_n
  \sinh \left[ \frac{2(\omega_n - i \rho_\sigma h) d}{\hbar v} \right]
      \nonumber \\
    &-& ( K^2 -1 - 2 Z_\theta^2 ) \Omega_n^2 \cos( 2 q d )
      + 2 Z_\theta
  \left( K^2 -1 -  Z_\theta^2 \right)^{1/2}  \Omega_n^2
     \sin( 2 q d )  ,
\end{eqnarray}
where
\begin{eqnarray}
  \label{kappa}
    K = \frac{1}{2} \left( \tilde{q} + \frac{1+Z_\theta^2}{ \tilde{q}}
        \right) .
\end{eqnarray}
We emphasize that the obtained expressions are consistent with
previous formulae for the Josephson current. For $h=0$, Eqs.
(\ref{mains})--(\ref{kappa}) are generalization of the
Furusaki-Tsukada formula\cite{Furusaki Tsukada 91} to
double-barrier  SINIS junctions with  $Z\ne 0$. For equal Fermi
energies of the two metals and for transparent interfaces,
$\kappa=1$ and $Z=0$, the rapidly oscillating terms are absent,
and Eq. (\ref{mains}) reduces to the  well known quasiclassical
expression in the clean limit.\cite{buzdin,zr} In Eq.
(\ref{gamman}), a weak exchange potential is taken into account
only by its contribution to the phase of the superconducting pair
potential, $-i \rho_\sigma h d / \hbar v$ in sinh and cosh terms,
that implies oscillations of $I(\phi)$ and changes the sign of the
current at the crossovers between $0$ and $\pi$ states. For $ k_B
T_c \ll h/E_F^{(F)} < 0.1$, the current-phase relation is almost a
universal function of the parameter $\Theta = (k_F^{(F)} d )
(h/E_F^{(F)})$, which measures the total influence of the
ferromagnet. For a stronger ferromagnet this is not the case, and
the general Eq. (\ref{main}) has to be applied. In all
illustrations (Figs. $1-6$) we have used Eq. (\ref{main}),
characterizing superconductors with $\Delta / E_F^{(S)} =
10^{-3}$.

Characteristic feature of the ballistic SIFIS junctions is an
oscillatory dependence of  $I(\phi)$ and  $I_c$ on $h$ and  $d$,
which is related to the crossovers between 0 and $\pi$ states.
However, even in SINIS junctions (where 0 state is the equilibrium
one) geometrical oscillations of the supercurrent occur due to the
coherent contribution of the quasiparticle transmission
resonances.\cite{brinkman,zetp} To stress this effect, in Fig. 1
we show an example of a thin and weak ferromagnet, $h/E_F^{(F)}
=0.01$, and compare it to a nonmagnetic-metal interlayer, $h=0$,
for the same interfacial transparency, $Z=1$, at low temperature
$T/T_c = 0.1$. In this case, geometrical oscillations are
dominant, the SIFIS junction being also in the 0 state (the first
crossover from 0 to $\pi$ state occurs for $d k_F^{(F)} = 125$).

The interplay between geometrical  oscillations and those induced
by a strong exchange potential is shown  in Fig. 2 for thin
ferromagnetic layers with  $h/E_F^{(F)} = 0.9$. Oscillations of
$I_c (d)$ due to the exchange potential are  shown in Fig. 2(a)
for a junction with transparent interfaces, $Z=0$, equal Fermi
energies, $\kappa=1$, and for two temperatures. For finite
interfacial transparency,  $Z=1$, these oscillations  are
superimposed on the geometrical ones, Fig. 2(b). In the same
figure, the influence of different band-widths in S and F metals
is also shown for $\kappa = 0.7$. Here, we use the normalization
$I_c R$, instead of $I_c R_N$ used in Fig. 1, to clearly show the
influence of the junction parameters ($h,~d,~Z$, and $\kappa$) on
the maximum supercurrent. Mean values of the normal resistance
corresponding, for example, to solid curves in Figs. 2(a) and 2(b)
are $R_N / R = 2.34$ and $4.55$, respectively.

The characteristic variation of nonsinusoidal $I(\phi)$ in the
vicinity of the crossover between  $0$ and $\pi$ states is
illustrated in  Fig. 3 for a highly resistive junction with the
same parameters used in Fig. 2(b), solid curve. Lower transparency
and FWVM, $Z=1$ and $\kappa=0.7$, shift the crossover point at
$T/T_c=0.1$ from $d_c = 9.45 / k_F^{(F)}$ in a transparent
junction (second dip of solid curve in Fig. 2(a)) to $d_c = 8.72 /
k_F^{(F)}$ (second dip of  solid curve in Fig. 2(b)). The region
of coexisting $0$ and $\pi$ states, $8.63 < d k_F^{(F)} < 8.82$,
is two times narrower than that in the transparent junction, $9.2
< d k_F^{(F)} < 9.6$. With increase of temperature or decrease of
transparency the contribution from the higher-order scattering
processes becomes negligible, transition regions become narrower
and $I(\phi)$ approaches the ordinary sinusoidal dependence, $\pm
I_c \sin\phi$, where $\pm$ correspond to 0 and $\pi$ junction,
respectively. Note that similar highly nonsinusoidal variation of
$I(\phi)$ also occurs in SF-FS Josephson junctions with
transparent geometrical constrictions.\cite{golubov1} However,
such a behavior is not stable against a disorder. In diffusive
double-barrier SIFIS junctions $I(\phi)$ does not cross the $\phi$
axis in the interval between $0$ and $\pi$.\cite{golubov1}

The temperature variation of $I_c$ is usually a monotonic decay
with increasing temperature. However, depending on parameters of
the junction, the transition between 0 and $\pi$ states can be
induced by changing the temperature. In that case, $I_c$ manifests
nonmonotonic dependence on temperature with a well-pronounced dip
at the transition. This is illustrated in Fig. 4 for the SIFIS
junctions with $h/E_F^{(F)} =0.92$, $Z=1.2$, $\kappa=1$, and three
values of $d$ close to $d_c=17.27/k_F^{(F)}$ at zero temperature
($d_c=17.14 / k_F^{(F)}$ at $T=0.9 T_c$). Three characteristic
$I_c (T)$ curves are shown for $ d k_F^{(F)} = 17$ (0 state),
$17.4$ ($\pi$ state), and $17.23$ (the temperature increase
induces a transition from 0 to $\pi$ state at $T/T_c = 0.22$). A
considerably large transition region of coexisting 0 and $\pi$
states, $0.1 < T/T_c < 0.3$, is shown in Fig. 5. Similar $0-\pi$
transitions occur at different temperatures in a very narrow
region $17.2 < d k_F^{(F)} < 17.3$ about the crossover $d_c$ at
zero temperature. We emphasize that the temperature-induced
transition takes place in the vicinity of any crossover point of
the junctions with finite transparency, with or without FWVM, and
with strong ferromagnetic influence. For example, the
temperature-induced transitions occur in the vicinity of the
crossover points in Fig. 2(b), represented by dips in both solid
and dotted $I_c (d)$ curves. However, this is not the case for
transparent interfaces, for example in the vicinity of dips in
solid $I_c (d)$ curve in Fig. 2(a). These results are in agreement
with the general conditions for the occurrence of the temperature
induced $0-\pi$ transition, given in Ref. \cite{barash}.

For thick layers of a weak ferromagnet,  $h/E_F^{(F)} = 0.01$,
oscillations of $I_c (d)$ due to the exchange potential are  shown
in Fig. 6(a) for a junction with transparent interfaces, $Z=0$,
equal Fermi energies, $\kappa=1$, and for two temperatures. The
influence of interfacial resistance and of FWVM at low temperature
is illustrated in  Fig. 6(b). One can  see that the contribution
of geometrical resonances is negligible in that case. Oscillations
of $I_c (d)$ in the   resistive junctions ($R_N / R = 5,~2.5$, and
$2.2$ for solid, dotted, and dashed curves in Fig. 6(b),
respectively)  are similar to those in the non-resistive one, Fig.
6(a), with shifted $d_c$ and significant lowering of amplitudes.
However, regions of coexisting  0 and $\pi$ states are
considerably narrower in the resistive junctions. For example, at
low temperature, $T/T_c = 0.1$, in a junction with transparent
interfaces and without FWVM, the crossover from 0  to $\pi$ state
occurs at $d_c k_F^{(F)}=111$ (first dip in solid curve, Fig.
6(a)) with coexisting  0 and   $\pi$ states in the region $60 < d
k_F^{(F)} < 130$, while in a junction with finite transparency and
FWVM ($Z=1$, $\kappa=0.7$), the corresponding crossover occurs at
$d_c k_F^{(F)}=120$ (first dip in solid curve, Fig. 6(b)), and the
coexisting  region, $110 < d k_F^{(F)} < 125$, is five times
narrower. Because of weak ferromagnetic influence the
temperature-induced transition is not found for the parameters
displayed in Fig. 6.

\section{Concluding remarks}

We have derived an  expression for the Josephson current in planar
ballistic SIFIS junctions, generalizing the Furusaki - Tsukada
formula\cite{Furusaki Tsukada 91} so that it includes interfacial
non-transparency and ferromagnetism in the normal-metal
interlayer. We used a non-self-consistent step function for the
pair potential, but in the case of low transparency, FWVM, and/or
thin interlayers, our results will not be altered significantly.
In that case, the depletion of the pair potential in the
superconductors is negligible. Characteristic proximity effects at
transparent FS interfaces have to be studied by a self-consistent
numerical treatment.\cite{halterman} In order to obtain simpler
expressions for the scattering amplitudes, we have neglected,
except in exponentials, the small energy terms in the wave
vectors, since their contribution is typically less than 0.1$\%$.
These terms are not neglected, however, in the exponentials, so
that we take into account the significant contribution from both
resonant and bound states, represented by rapidly and slowly
oscillating terms.

The solutions obtained for the Andreev and the normal reflection
amplitudes provide a fully microscopic study of the coherent
superconducting properties in  ballistic double-barrier junctions
with ferromagnetic, or nonmagnetic normal-metal interlayer. The
resulting wave functions and the quasiparticle excitation energies
can be used to compute all physically relevant quantities, e.g.
the local DOS and the superconducting pair amplitude.\cite{foot2}
These applications of our results are left for future work.

Qualitatively, our results confirm previously obtained main
features of the metallic SFS systems, and uncover new coherency
effects in nanostructured ballistic junctions. The pronounced
geometrical  oscillations of the supercurrent occur in
double-barrier SINIS junctions with thin interlayers of a clean
normal metal, due  to the coherent contribution of the
quasiparticle transmission resonances to the Andreev bound
states.\cite{zetp} The amplitudes of the supercurrent oscillations
are significantly larger than those of the normal current in the
corresponding NININ ballistic junction. For thin layer of a strong
ferromagnet, we found that geometrical oscillations are
superimposed on the oscillations induced by  crossovers between 0
and $\pi$ states. For high interfacial transparency and/or  thick
interlayers, coherency effects are less pronounced, in agreement
with previous theoretical results. \cite{brinkman} Low interfacial
transparency and FWVM affect the position of crossover points, and
narrower the transition regions of coexisting 0 and $\pi$ states.

We have shown that the temperature induced transition occurs in
ballistic  SIFIS junctions with finite interfacial transparency
and  strong ferromagnetic influence,  if  the parameters of the
junction are  sufficiently close to the crossover  at zero
temperature. The characteristic  nonmonotonic variation of the
maximum Josephson current with temperature  agrees  with previous
experimental and theoretical results.
\cite{ryazanov,fogelstrom,barash,chtchelkatchev,krivoruchko,bergeret,golubov,barash1,zyuzin}
However, in the ballistic junctions the transition region of
coexisting 0 and $\pi$ states is considerably large. This effect
can be exploited, for example, in the design of a $\pi$ SQUID with
improved accuracy, which operates as a usual one with effectively
two times smaller flux quantum.\cite{zr} Such a device has
potential applications in novel quantum electronics.\cite{blatter}

\section{Acknowledgments
} The work was supported in part by  the Greek-Yugoslav Scientific
and Technical Cooperation  program on Superconducting
Heterostructures and  Devices. Z. R. acknowledges also the support
of the Serbian Ministry of Science, Technology and Development,
Project No. 1899, and thanks Ivan Bo\v zovi\'c and Milo\v s Bo\v
zovi\'c for useful discussions.

\begin{figure}[h]
\centerline{\hbox{
  \psfig{figure=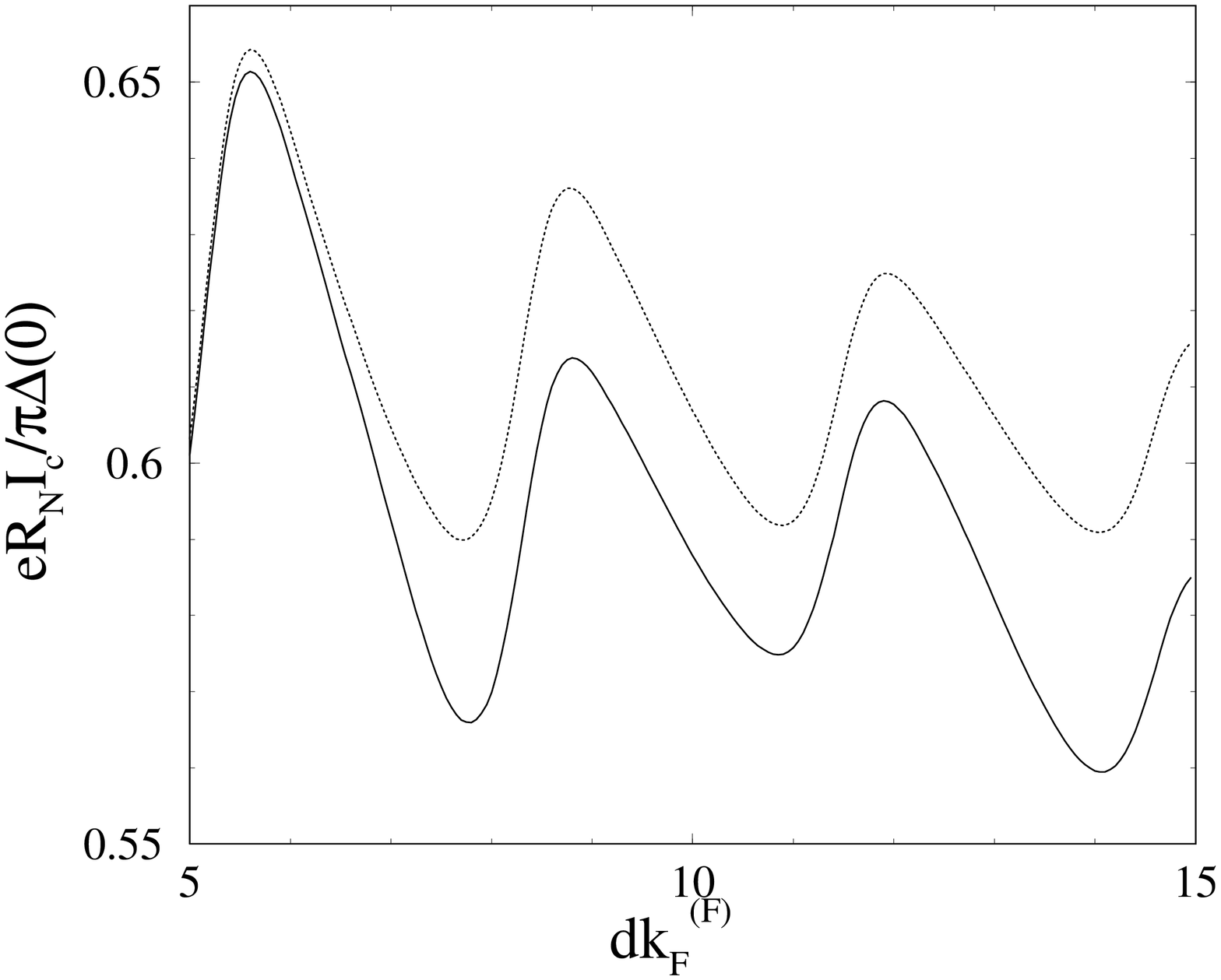,height=80mm,width=80mm,angle=-90}
}}
  \caption{
  Maximum current $I_c$
  as a function of $d$
  for $T/T_c = 0.1$, $\kappa=1$,  $Z=1$,  and for
  $h/E_F^{(F)}=0.01$ (solid curve), and for
  $h=0$ (dotted curve). In all illustrations the superconductors
  are characterized by $\Delta/E_F^{(S)}=10^{-3}$.
}
\end{figure}

\vspace{10cm}

\begin{figure}[h]
\centerline{\hbox{
  \psfig{figure=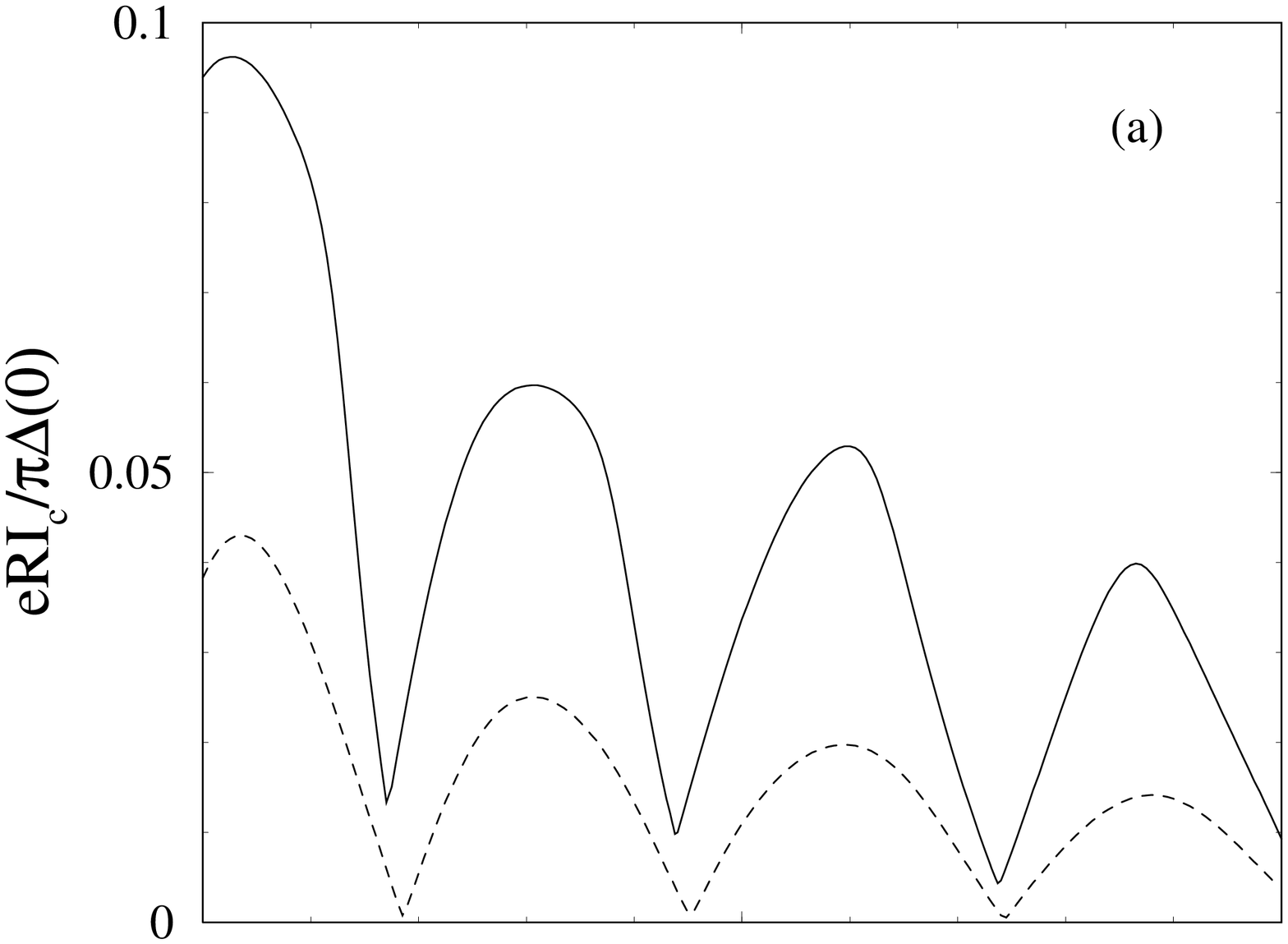,height=75mm,width=78mm,angle=-90}
}}
\centerline{\hbox{
  \psfig{figure=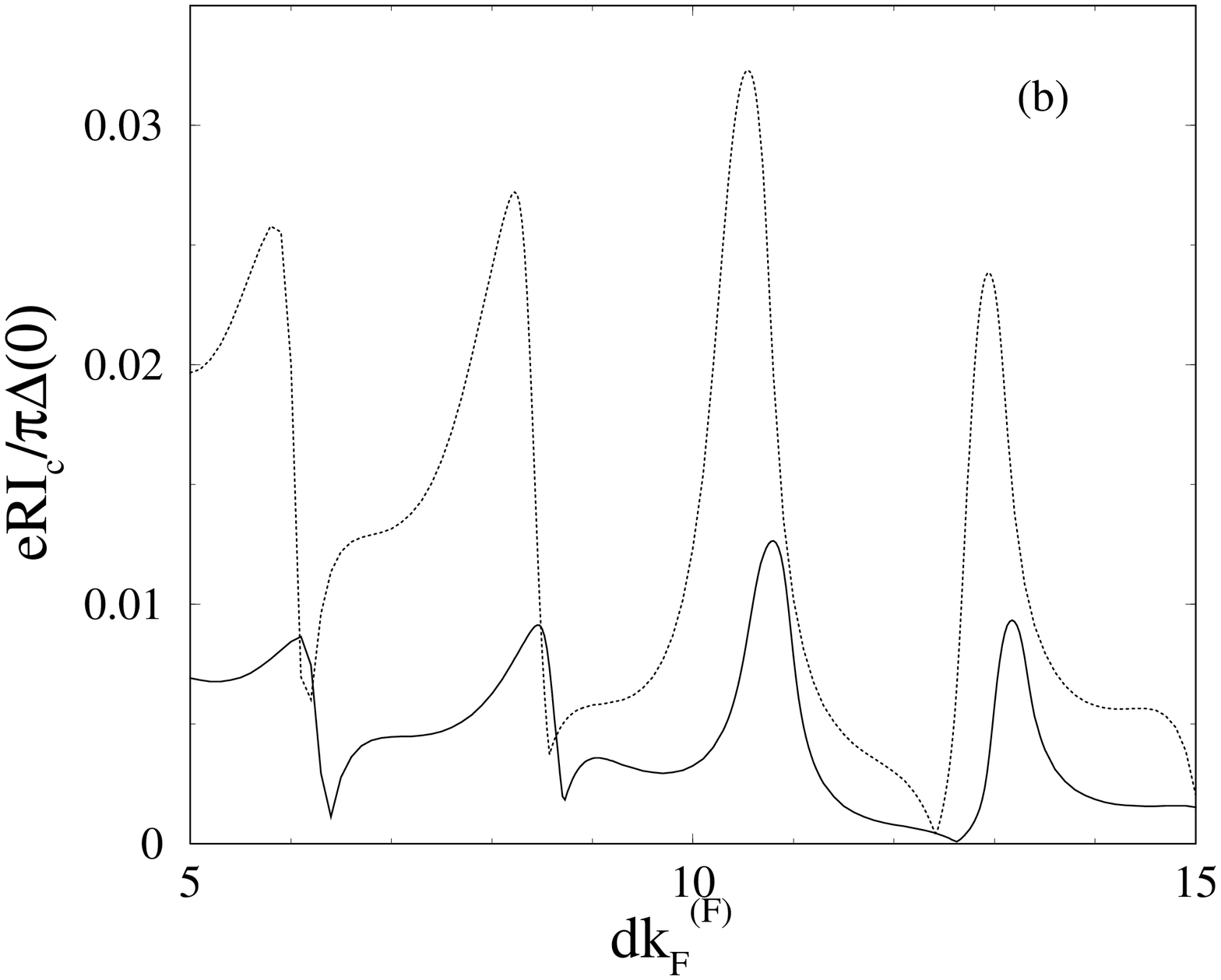,height=80mm,width=80mm,angle=-90}
}}
  \caption{
   Maximum  current $I_c$
  as a function of $d $
  for $h/E_F^{(F)}=0.9$:
  (a) $Z=0$, $\kappa=1$,
      $T/T_c = 0.1$ (solid curve) and  $T/T_c = 0.7$
  (dashed curve);
  (b) $T/T_c = 0.1$, and $Z=1$, $\kappa=0.7$ (solid curve),
     $Z=1$, $\kappa=1$ (dotted curve).
  Dips  in  $I_c (d)$ separate alternating
  0 and $\pi$ states, starting with 0 state  from the left.
}
\end{figure}

\begin{figure}[h]
\centerline{\hbox{
  \psfig{figure=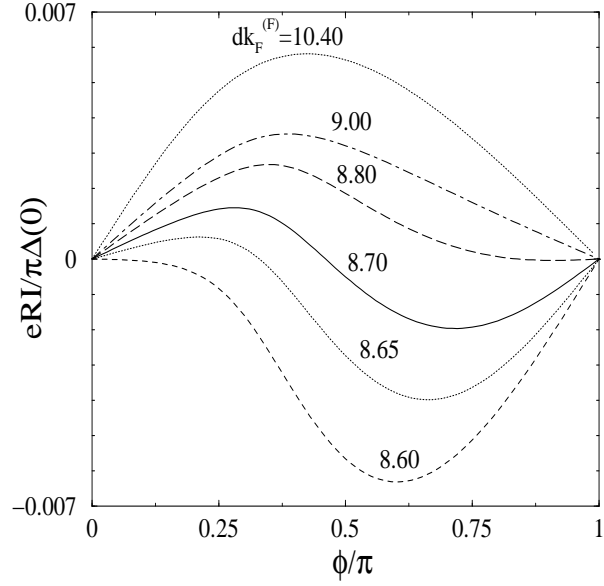,height=80mm,width=80mm,angle=-90}
}}
  \caption{
  Current-phase relation, $I(\phi)$,
  for $T/T_c = 0.1$, $h/E_F^{(F)}=0.9$,
  $Z=1$, $\kappa=0.7$, and for five values of $d k_F^{(F)}$
  in the vicinity of the crossover  between $0$ and  $\pi$  states
  ($d_c k_F^{(F)} = 8.72$), see solid curve in Fig. 2(b).
}
\end{figure}

\begin{figure}[h]
\centerline{\hbox{
  \psfig{figure=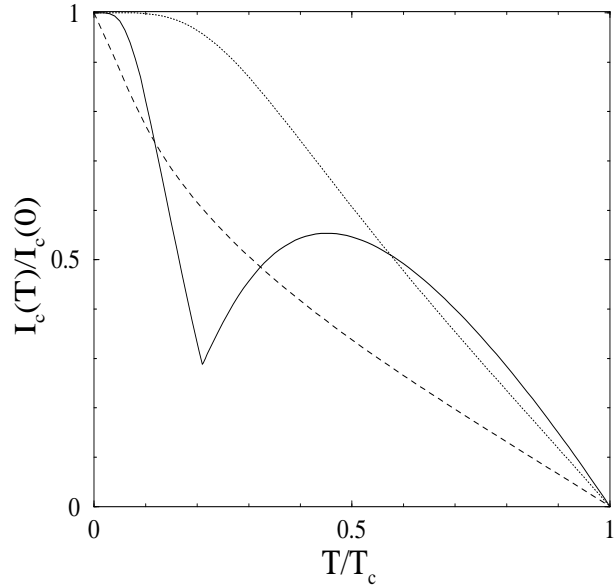,height=80mm,width=80mm,angle=-90}
}}
  \caption{
   Temperature variation of  $I_c (T)$,
  normalized by $I_c (0)$, for $h/E_F^{(F)}=0.92$, $Z=1.2$, $\kappa=1$,
  and for three values of $d k_F^{(F)}=17$ (dotted curve),
  $17.23$ (solid curve) and $17.4$  (dashed curve).
}
\end{figure}

\begin{figure}[h]
\centerline{\hbox{
  \psfig{figure=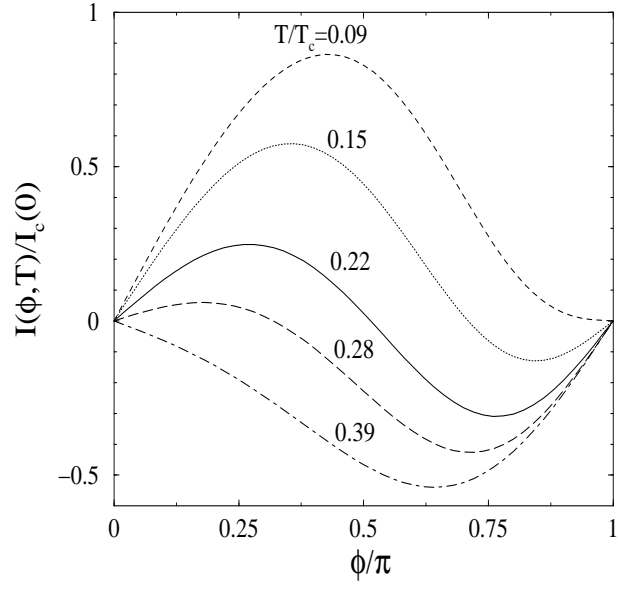,height=80mm,width=80mm,angle=-90}
}}
  \caption{
   Temperature variation of  $I(\phi,T)$,
   normalized by $I_c (0)$, for $h/E_F^{(F)}=0.92$, $Z=1.2$, $\kappa=1$,
   $d k_F^{(F)}=17.23$ and for five values of $T/T_c$
   in the vicinity of the transition from 0 to $\pi$  state,
   see solid curve in Fig. 4.
}
\end{figure}

\vspace{10cm}

\begin{figure}[h]
\centerline{\hbox{
  \psfig{figure=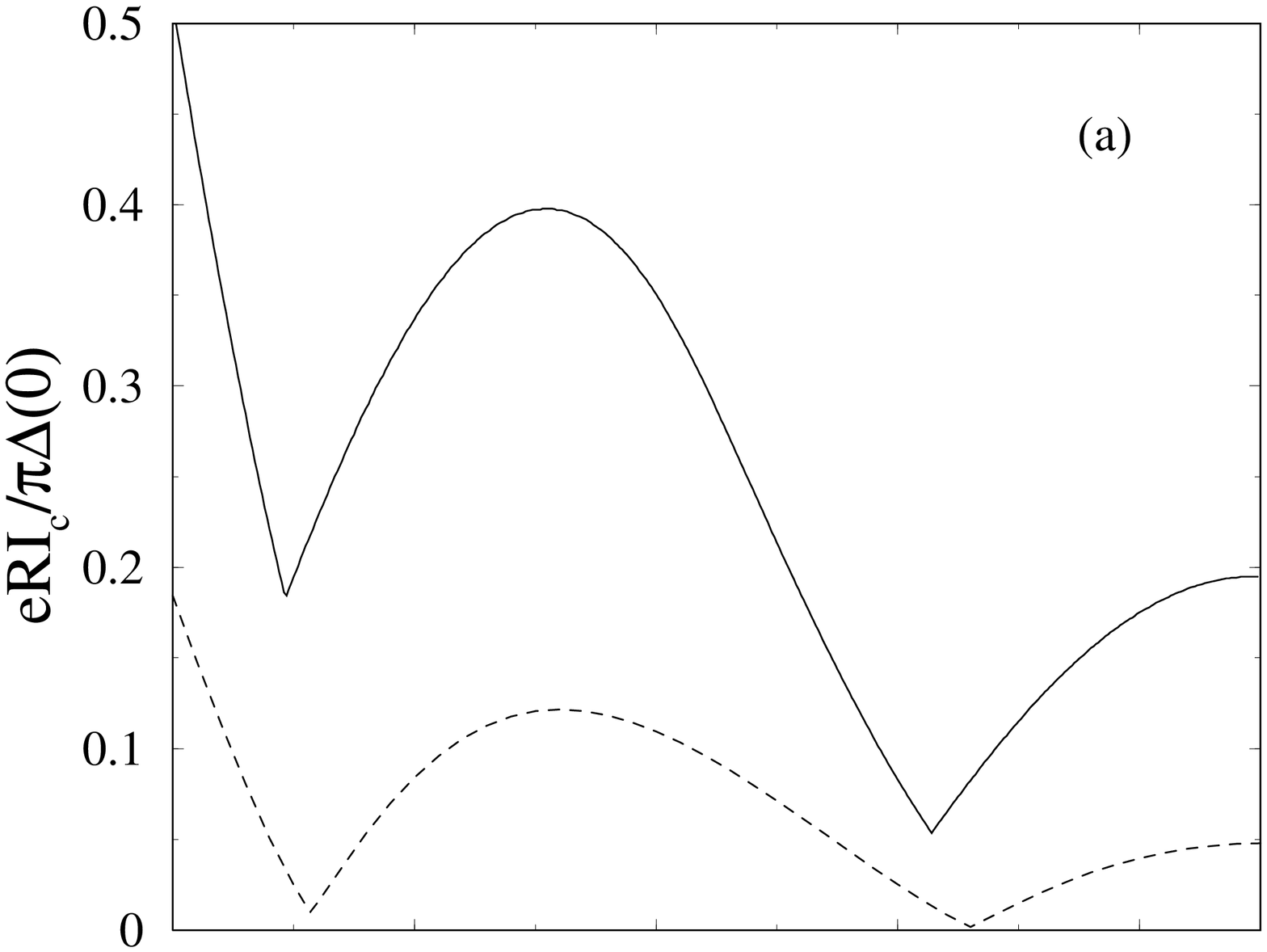,height=75mm,width=80mm,angle=-90}
}}
\centerline{\hbox{
  \psfig{figure=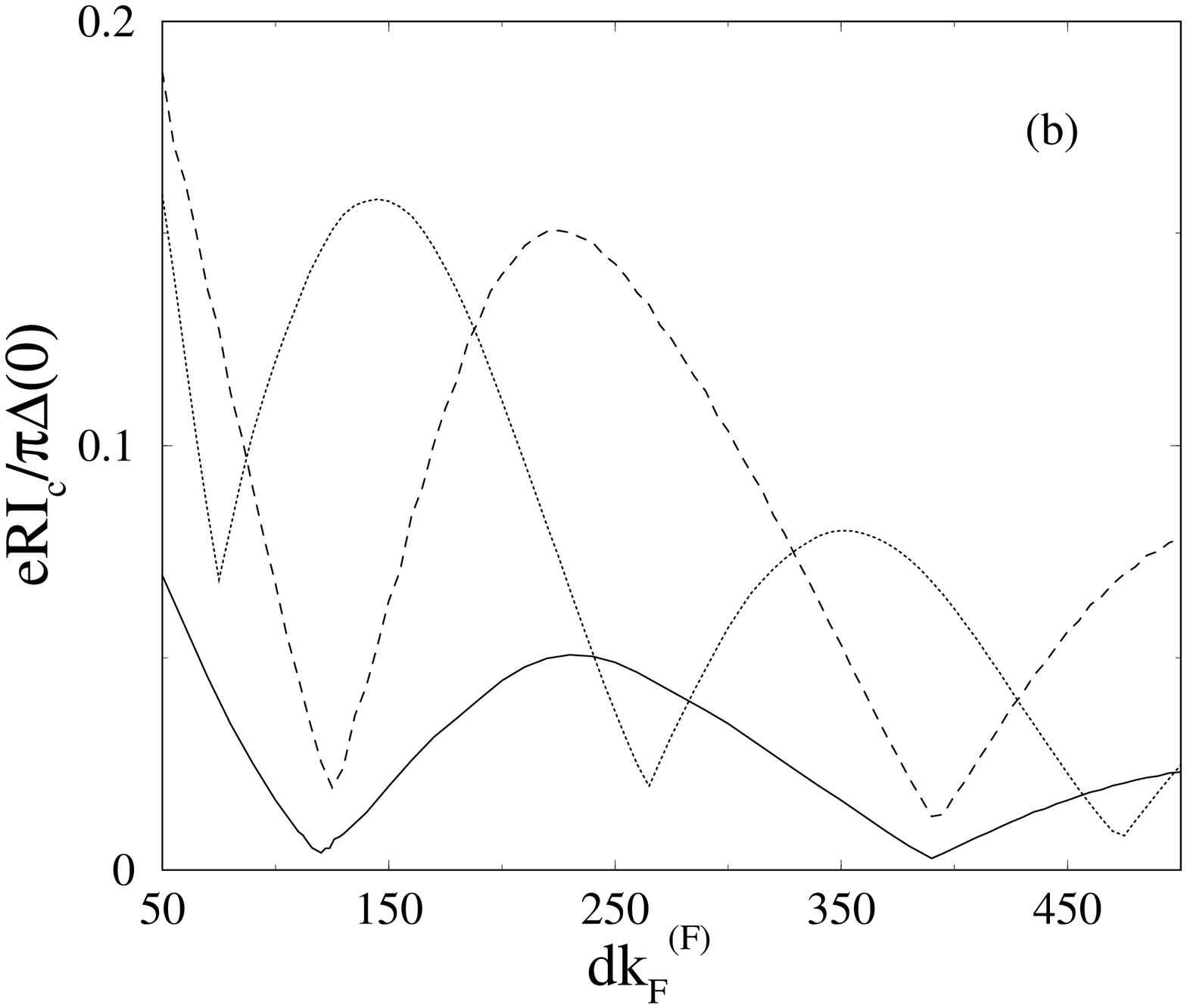,height=80mm,width=80mm,angle=-90}
}}
  \caption{
   Maximum  current $I_c$
  as a function of $d $
  for $h/E_F^{(F)}=0.01$:
  (a)  $Z=0$, $\kappa=1$,
  $T/T_c = 0.1$ (solid curve) and  $T/T_c = 0.7$ (dashed curve);
  (b) $T/T_c = 0.1$, and $Z=1$, $\kappa=0.7$ (solid curve),
  $Z=1$, $\kappa=1$ (dashed curve),
  $Z=0$, $\kappa=0.7$ (dotted curve).
  Dips  in  $I_c (d)$ separate alternating
  0 and $\pi$ states, starting with 0 state  from the left.
}
\end{figure}


\begin{thebibliography}{99}

\bibitem{tedrow}
    P. M. Tedrow and R. Meservey,
    Phys. Rep. {\bf  238}, 173 (1994).

\bibitem{ryazanov}
   V. V. Ryazanov, V. A. Oboznov, A. Yu. Rusanov,
   A. V. Veretennikov, A. A. Golubov, and J. Aarts,
    Phys. Rev.  Lett. {\bf 86}, 2427  (2001).

\bibitem{ryazanov1}
   V. V. Ryazanov, V. A. Oboznov,
   A. V. Veretennikov, and A. Yu. Rusanov,
   Phys. Rev. B {\bf 65}, 020501 (2001).

\bibitem{kontos}
    T. Kontos, M. Aprili, J. Lesueur, and X. Grison,
    Phys. Rev.  Lett. {\bf 86}, 304 (2001).

\bibitem{geers}
     J. M. E. Geers, M. B. S. Hesselberth, J. Aarts,
     and A. A. Golubov,
     Phys. Rev. B {\bf 64}, 094506 (2001).

\bibitem{bourgeois}
    O. Bourgeois, P. Gandit, J. Lesueur, A. Sulpice, X. Grison,
    and J. Chaussy,
    Eur. Phys. J. B {\bf  21}, 75 (2001).

\bibitem{kontos1}
    T. Kontos, M. Aprili, J. Lesueur, F. Gen\^{e}t,
    B. Stephanidis, and R. Boursier,
    Phys. Rev.  Lett. {\bf 89}, 137007 (2002).

\bibitem{fogelstrom}
    M. Fogelstr\"om,
    Phys. Rev. B {\bf 62}, 11812 (2000).

\bibitem{barash}
    Yu. S. Barash and I. V. Bobkova,
    Phys. Rev. B {\bf 65}, 144502 (2002).

\bibitem{chtchelkatchev}
    N. M. Chtchelkatchev, W. Belzig, Yu. V. Nazarov, and C. Bruder,
    Pis'ma Zh. \'Eksp. Teor. Fiz. {\bf 74}, 357  (2001)
    [JETP Lett. {\bf 74}, 323 (2001)].

\bibitem{krivoruchko}
    V. N. Krivoruchko and E. A. Koshina,
    Phys. Rev. B {\bf 64}, 172511 (2001).

\bibitem{bergeret}
     F. S. Bergeret, A. F. Volkov, and K. B. Efetov,
     Phys. Rev. B {\bf 64}, 134506 (2001).

\bibitem{golubov}
     A. A. Golubov, M. Yu. Kupriyanov, and Ya. V. Fominov,
     Pis'ma Zh. \'Eksp. Teor. Fiz. {\bf 75}, 223  (2002)
     [JETP Lett. {\bf 75}, 190 (2002)].

\bibitem{barash1}
    Yu. S. Barash, I. V. Bobkova, and T. Kopp,
    Phys. Rev. B {\bf 66}, 140503 (2002).

\bibitem{zyuzin}
    A. Yu. Zyuzin, B. Spivak, and M. Hru\v{s}ka,
    Europhys. Lett. {\bf  62}, 97 (2003).

\bibitem{zr}
   Z. Radovi\'c, L. Dobrosavljevi\'c-Gruji\'c, and B. Vuji\v ci\'c,
     Phys. Rev. B {\bf 63}, 214512  (2001).

\bibitem{golubov1}
     A. A. Golubov, M. Yu. Kupriyanov, and Ya. V. Fominov,
     Pis'ma Zh. \'Eksp. Teor. Fiz. {\bf 75}, 588  (2002)
     [JETP Lett. {\bf 75}, 709 (2002)].

\bibitem{heikkila}
    T. T. Heikkil\"{a}, F. K. Wilhelm, and G. Sch\"{o}n,
   Europhys. Lett. {\bf  51}, 434 (2000).

\bibitem{halterman}
    K. Halterman and O. T. Valls,
    Phys. Rev. B {\bf 65}, 014509 (2001); {\bf 66}, 224516 (2002).

\bibitem{zareyan}
    M. Zareyan, W. Belzig, and Yu. V. Nazarov,
    Phys. Rev.  Lett. {\bf 86}, 308 (2001).

\bibitem{baladie}
     I. Baladie  and A. Buzdin,
     Phys. Rev. B {\bf 64}, 224514 (2001).

\bibitem{bergeret1}
     F. S. Bergeret, A. F. Volkov, and K. B. Efetov,
     Phys. Rev. B {\bf 65}, 134505 (2002).

\bibitem{brinkman}
    A. Brinkman and A. A. Golubov,
    Phys. Rev. B {\bf 61},  11297 (2000).

\bibitem{ingerman}
    A. Ingerman, G. Johansson, V. S. Shumeiko, and G. Wendin,
    Phys. Rev. B {\bf 64}, 144504 (2001).

\bibitem{kikuchi}
     K. Kikuchi, H. Imamura, S. Takahashi, and S. Maekawa,
     Phys. Rev. B {\bf 65}, 020508 (2001).

\bibitem{bozovic}
     M. Bo\v{z}ovi\'{c} and Z. Radovi\'{c},
     Phys. Rev. B {\bf 66}, 134524 (2002).

\bibitem{nevirkovets}
    I. P. Nevirkovets, J. B. Ketterson, and S. Lomatch,
    Appl. Phys. Lett.  {\bf 74}, 1624 (1999).

\bibitem{schulze}
    H. Schulze, R. Behr, F. M\"uller, and J. Niemeyer,
    Appl. Phys. Lett.  {\bf 73}, 996 (1998).

\bibitem{moussy}
    N. Moussy, H. Courtois, and B. Pannetier,
    Rev. Sci. Instrum. {\bf 72}, 128 (2002).

\bibitem{bulaevskii}
     L. N. Bulaevskii, V. V. Kuzii, and A. A. Sobyanin,
     Pis'ma Zh. \'Eksp. Teor. Fiz. {\bf 25}, 314  (1977)
    [JETP Lett. {\bf 25}, 290  (1977)].

\bibitem{buzdin}
    A. I. Buzdin, L. N. Bulaevskii, and S. V. Paniukov,
     Pis'ma Zh. \'Eksp. Teor. Fiz. {\bf 35}, 147  (1982)
     [JETP Lett. {\bf 35}, 178  (1982)].

\bibitem{radovic}
     Z. Radovi\'c, M. Ledvij, Lj. Dobrosavljevi\'c-Gruji\'c,
     A. I. Buzdin, and J. R. Clem,
     Phys. Rev. B {\bf 44}, 759 (1991).

\bibitem{buzdin1}
     A. I. Buzdin and M. V. Kupriyanov,
      Pis'ma Zh. \'Eksp. Teor. Fiz. {\bf 52}, 1089,  (1990)
      [JETP Lett. {\bf 52}, 487  (1990)].

\bibitem{jiang}
     J. S. Jiang, D. Davidovi\'c, D. H. Reich, and C. L. Chien,
     Phys. Rev.  Lett. {\bf 74}, 314 (1995).

\bibitem{obi}
     Y. Obi, M. Ikebe, T. Kubo, and H. Fujimori,
     Physica C {\bf 317-318}, 149 (1999).

\bibitem{vanharlingen}
     D. J. van Harlingen,
     Rev. Mod. Phys. {\bf 67}, 515 (1995).

\bibitem{baselmans}
     J. J. A. Baselmans, T. T. Heikkil\"{a}, B. J. van Wees,
     and T. M. Klapwijk,
     Phys. Rev. Lett. {\bf 89}, 207002 (2002);
     J. J. A. Baselmans, A. F. Morpurgo, B. J. van Wees, and
     T. M. Klapwijk, Nature (London) {\bf 397}, 43 (1999).


\bibitem{andreev}
     A. F. Andreev,
     Zh. Eksp. Teor. Fiz. {\bf 46}, 1823 (1964)
    [Sov. Phys. JETP {\bf 19}, 1228 (1964)].

\bibitem{BTK}
     G. E. Blonder, M. Tinkham, and T. M. Klapwijk,
     Phys. Rev. B {\bf 25}, 4515 (1982).

\bibitem{Furusaki Tsukada}
     A. Furusaki and M. Tsukada,
      Solid State  Commun. {\bf 78}, 299 (1991).

\bibitem{beenakker}
     C. W. J. Beenakker,
     Phys. Rev.  Lett. {\bf 67}, 3836 (1991).

\bibitem{dejong}
    M. J. M. de Jong and C. W. J. Beenakker,
    Phys. Rev.  Lett. {\bf 74}, 1657 (1995).

\bibitem{Tanaka 97}
    Y. Tanaka and S. Kashiwaya,
    Physica {\bf C 274}, 357  (1997).

\bibitem{Tanaka 10}
    Y. Tanaka and S. Kashiwaya,
    J. Phys. Soc. Japan {\bf 69}, 1152 (2000).

\bibitem{beasley}
    S. Kashiwaya, Y. Tanaka, N. Yoshida, and M. R. Beasley,
      Phys. Rev. B {\bf 60}, 3572  (1999).

\bibitem{Furusaki Tsukada 91}
     A. Furusaki and M. Tsukada,
      Phys. Rev. B {\bf 43}, 10164  (1991).

\bibitem{arnold}
    G. B. Arnold,
     Phys. Rev. B {\bf 18}, 1076 (1978).

\bibitem{zetp}
     A. L. Gudkov, M. Y. Kupriyanov, and K. K. Likharev,
     Zh. Eksp. Teor. Fiz. {\bf 94}, 319 (1988)
    [Sov. Phys. JETP {\bf 67}, 1478 (1988)].


\bibitem{foot2}
The matrix Green's functions for both superconductors and
normal-metal interlayer can be expressed  through the reflection
amplitudes $a_1,~a_2,~b_1$, and $b_2$, see Refs. \cite{Furusaki
Tsukada,Furusaki Tsukada 91}.

\bibitem{blatter}
   G. Blatter, V. B. Geshkenbein, and L. B. Ioffe,
  Phys. Rev. B {\bf 63}, 174511 (2001);
  L. B. Ioffe, V. B. Geshkenbein, M. V. Feigelman,
  A. L. Fauchere,  and G. Blatter,
  Nature (London) {\bf 398}, 679 (1999).

\end{thebibliography}
\end{document}